     \documentstyle[11pt]{article}
\font\titlfnt = cmss17

\begin{document}

\bibliographystyle{unsrt}

\renewcommand{\thesection}{\Roman{section}} 
\renewcommand{\thefootnote}{\fnsymbol{footnote}} 

\begin{titlepage}

\rightline{hep-th/9606164}\par

\vspace{7.5mm}

\begin{center}

{\titlfnt
Non-Grassmann ``Classicization'' of Fermion Dynamics}

\vspace{10mm}

\noindent
{\large
S. K. Kauffmann}\footnote{
E-mail: skk@sydney.dialix.oz.au;
Fax: +61 2 99992412; Tel: +61 2 99796301}\\
Unit 3 / 51-53 Darley Street, Mona Vale, NSW 2103, Australia.

\vspace{5mm}

\end{center}

\vspace{10mm} 

\begin{abstract}
A carefully motivated symmetric variant of
the Poisson bracket in ordinary (not
Grassmann) phase space variables is shown
to satisfy identities which are in
algebraic correspondence with the
anticommutation postulates for quantized
Fermion systems.  ``Symplecticity'' in
terms of this symmetric Poisson bracket
implies generalized Hamilton's equations
that can only be of Schr\"odinger type
(e.g., the Dirac equation but not the
Klein-Gordon or Maxwell equations).
This restriction also excludes the old
``four-Fermion'' theory of beta decay.

\vspace{5mm}

\vspace{5mm}
\end{abstract}

\vspace{85mm}

\noindent
June 1996

\end{titlepage}

\renewcommand{\thefootnote}{\arabic{footnote}} 
\addtocounter{footnote}{-\value{footnote}} 


\addtocounter{equation}{-\value{equation}}

Quantized Fermion dynamics, with its Pauli exclusion
principle, no more possesses a limit of ``large''
quantum numbers than does elementary spin one-half
quantum dynamics.  Thus the notion of
``classicizing'' Fermion dynamics via a
formal $\hbar\to 0$ limit \cite{Casal1} is physically
unsound.  In fact, the ostensible Fermion
``classicization'' developed in
Refs.~\cite{Casal1,Casal2} maneuvers shy of this
trap by declining clean abandonment of quantum
noncommutativity, which lingers on in the
guise of anticommuting Grassmann phase space
variables (the oxymoronic tag
``anticommuting $c$-numbers'' notwithstanding).

The physically mismotivated (by the $\hbar\to 0$
notion) and theoretically ambiguous (definitely
not classical, but neither fully-fledged quantum)
Grassmann phase space variables are avoided here
in favor of true $c$-number phase space variables,
which are used to construct a heuristically compelling
``symmetric'' variant of the Poisson bracket (its
definition specifically requires that $\hbar\ne 0$).
This ``symmetric'' Poisson  bracket satisfies
phase-space vector component identities whose
algebraic relation to the postulated
{\em anticommutation\/} rules of quantized Fermion
dynamics fully parallels the algebraic relation of
the ordinary Poisson bracket phase-space vector
component identities to the postulated
{\em commutation\/} rules of quantized point particle
and Boson dynamics.

Given this soundly based ``symmetric'' Poisson bracket,
the structure of ``classical''
Fermion dynamics follows straightforwardly from
the requirements of ``symplecticity'' with respect
to it---the derivations can be carried out
in perfect parallel with the well-known ones of
ordinary (or Boson) classical dynamics \cite{Guillemin}.
For continuous one-parameter sequences of infinitesimal
Fermion ``canonical'' transformations, one obtains the
same natural generalization of Hamilton's equations as
occurs in ordinary classical dynamics \cite{Guillemin},
but one also finds stringent constraints on the form of
the ``generalized Hamiltonian functions'' or ``canonical
transformation generators'' which are permitted to appear
in these Fermion ``classical'' dynamical equations.
Indeed, the restrictions on these generators are such that
``classical'' Fermion dynamics must be {\em linear\/} and
described by a {\em Schr\"odinger\/} type of
equation (which may possibly be inhomogeneous).  The Dirac
equation, which is of Schr\"odinger type, can describe a
``classical'' Fermion system, but the inherently
non-Schr\"odinger (even though linear) Klein-Gordon and
Maxwell equations cannot.  Also the old ``four-Fermion''
theory of beta decay cannot describe a legitimate
``classical'' Fermion system (it is not thus forbidden
under the Grassmann variable regime).

Ordinary classical dynamics is usually discussed in terms
of real-valued phase space vector variables of the form
$(\vec q,\vec p)$.  However, its relation to the quantum
theory and to Fermion systems is much more transparent if
one changes these real phase space vector variables to the
complex-valued dimensionless phase space vector variables
$\vec a \equiv (\vec q/q_s + iq_s\vec p/\hbar)/\sqrt{2}$
and their complex conjugates
$\vec a^{\, *} = (\vec q/q_s - iq_s\vec p/\hbar)/\sqrt{2}$,
where $q_s$ is a nonzero real-valued scale factor that has
the same dimensions as the components of $\vec q$ (note
also the obvious requirement that $\hbar\ne 0$).  In terms
of $\vec a$ and $\vec a^{\, *}$, $(\vec q,\vec p) =
(q_s(\vec a + \vec a^{\, *}),-i\hbar (\vec a - \vec a^{\, *})/q_s)/
\sqrt{2}$.  In terms of components of both of these types
of phase space vector variable, the usual Poisson bracket of
ordinary classical dynamics is
\begin{equation}
\{f,\, g\} \equiv \sum_k\left ({\partial f\over\partial q_k}
{\partial g\over\partial p_k} - {\partial g\over\partial q_k}
{\partial f\over\partial p_k}\right ) = -\,{i\over\hbar}
\sum_k\left ({\partial f\over\partial a_k}
{\partial g\over\partial a^*_k} - {\partial g\over\partial a_k}
{\partial f\over\partial a^*_k}\right ).
\label{Eq:Poisbr}
\end{equation}
{}From the second Poisson bracket representation given in
Eq.~(\ref{Eq:Poisbr}) above we abstract the ``sub-bracket''
\begin{equation}
\{f\circ g\} \equiv \sum_k{\partial f\over\partial a_k}
{\partial g\over\partial a^*_k}\, ,
\label{Eq:orPoisbr}
\end{equation}
which we call the {\em ordered\/} Poisson bracket.  We note
that while $\{f\circ g\}$ is linear in each of its two argument
functions $f$ and $g$, it is neither antisymmetric nor symmetric
(commutative) under their interchange.  However, it {\em does\/}
satisfy the identity $\{f\circ g\} = \{g^*\circ f^*\}^*$, which
is in algebraic correspondence with the Hermitian conjugation
formula for the {\em product of two Hilbert-space operators},
i.e., $\hat f\hat g = (\hat g^{\dagger}\hat f^{\dagger}
)^{\dagger}$.  This together with the fact that
$\{f,\, g\} = -i(\{f\circ g\} - \{g\circ f\})/\hbar$, as follows
from Eqs.~(\ref{Eq:Poisbr}) and (\ref{Eq:orPoisbr}), is a strong
heuristic motivation for the usual quantum theoretic postulates
that identify certain quantum operator commutators
$\hat f\hat g - \hat g\hat f$ with the corresponding Poisson
bracket expressions $i\hbar\{f,\, g\}$.  The factor of
$i\hbar$ which is involved can be eliminated by identifying these
commutators directly with the corresponding {\em antisymmetric\/}
Poisson brackets $\{f,\, g\}_- \equiv \{f\circ g\} - \{g\circ f\}$.
As natural counterparts to these one has the {\em symmetric\/}
Poisson brackets $\{f,\, g\}_+ \equiv \{f\circ g\} + \{g\circ f\}$,
which are the obvious ``classical'' candidates to correspond to
certain quantum operator {\em anticommutators\/} $\hat f\hat g +
\hat g\hat f$, such as those which enter into the quantum
postulates for Fermion systems.  Bearing in mind that
\begin{equation}
\{f,\, g\}_{\pm} =
\sum_k\left ({\partial f\over\partial a_k}
{\partial g\over\partial a^*_k} \pm
{\partial g\over\partial a_k}
{\partial f\over\partial a^*_k}\right ),
\label{Eq:asPoisbr}
\end{equation}
we readily calculate the symmetric and antisymmetric Poisson
brackets for the components of $\vec a$ and $\vec a^{\, *}$:
\begin{equation}
\{a_i,\, a_j\}_{\pm} = 0 = \{a_i^*,\, a_j^*\}_{\pm}, \quad
\{a_i,\, a_j^*\}_{\pm} = \delta_{ij} = \pm\{a_j^*,\, a_i\}_{\pm}.
\label{Eq:asPoisrl}
\end{equation}
The quantum commutation and anticommutation relations which would
algebraically correspond to Eqs.~(\ref{Eq:asPoisrl}) are:
\begin{equation}
\hat a_i\hat a_j\pm\hat a_j\hat a_i = 0 =
\hat a_i^{\dagger}\hat a_j^{\dagger}\pm
\hat a_j^{\dagger}\hat a_i^{\dagger}, \quad
\hat a_i\hat a_j^{\dagger}\pm
\hat a_j^{\dagger}\hat a_i = \delta_{ij}\hat I =
\pm(\hat a_j^{\dagger}\hat a_i\pm
\hat a_i\hat a_j^{\dagger}).
\label{Eq:asComrl}
\end{equation}
When $\pm = -$, we recognize Eqs.~(\ref{Eq:asComrl}) as the commutation
relations of the ladder operators for independent quantum harmonic
oscillators, while when $\pm = +$, we recognize Eqs.~(\ref{Eq:asComrl})
as the anticommutation relations of the creation and annihilation
operators for independent quantum Fermion system particle occupation
states.

The canonical transformations of ordinary classical dynamics
are mappings of the complex phase space vectors $\vec a\to\vec
A(\vec a,\vec a^{\, *})$ and
$\vec a^{\, *}\to (\vec A(\vec a,\vec a^{\, *}))^*$
which preserve the {\em antisymmetric\/} Poisson bracket relations
among the complex phase space vector components that are given by
Eqs.~(\ref{Eq:asPoisrl}) with $\pm = -$.  In view of the algebraic
correspondence with quantum Fermion systems established above, we
may confidently define the the canonical transformations of
Fermion system ``classical'' dynamics as those complex vector phase
space mappings which preserve the {\em symmetric\/} Poisson bracket
relations among the complex phase space vector components that are
given by Eqs.~(\ref{Eq:asPoisrl}) with $\pm = +$.

Specializing now to infinitesimal phase space transformations
$\vec a\to\vec A = \vec a + \delta\vec a(\vec a,\vec a^{\, *})$ in
the manner of Guillemin and Sternberg \cite{Guillemin}, we
readily calculate the antisymmetric and symmetric Poisson
brackets for the components of $\vec A$ and $\vec A^*$ to first
order in $\delta\vec a$ and $\delta\vec a^{\, *}$ from
Eq.~(\ref{Eq:asPoisbr}):
\[
\{A_i,\, A_j\}_{\pm} =
{\partial (\delta a_j)\over\partial a_i^*}\pm
{\partial (\delta a_i)\over\partial a_j^*}\, , \quad
\{A_i^*,\, A_j^*\}_{\pm} =
{\partial (\delta a_i^*)\over\partial a_j}\pm
{\partial (\delta a_j^*)\over\partial a_i}\, ,
\]
\begin{equation}
\{A_i,\, A_j^*\}_{\pm} = \delta_{ij} +
{\partial (\delta a_i)\over\partial a_j} +
{\partial (\delta a_j^*)\over\partial a_i^*} =
\pm\{A_j^*,\, A_i\}_{\pm}.
\label{Eq:inPoisbr}
\end{equation}
If we now impose the requirement that this infinitesimal phase
space vector transformation is {\em canonical\/} (i.e., that it
preserves the antisymmetric or symmetric Poisson bracket
relations among the complex phase space vector components given
by Eqs.~(\ref{Eq:asPoisrl})), we obtain the three equations:
\begin{equation}
{\partial (\delta a_j)\over\partial a_i^*} = \mp
{\partial (\delta a_i)\over\partial a_j^*}\, , \quad
{\partial (\delta a_j^*)\over\partial a_i}= \mp
{\partial (\delta a_i^*)\over\partial a_j}\, , \quad
{\partial (\delta a_i)\over\partial a_j} +
{\partial (\delta a_j^*)\over\partial a_i^*} = 0.
\label{Eq:inCan}
\end{equation}
The last of Eqs.~(\ref{Eq:inCan}) is independent of the value of
the $\mp$ symbol (i.e., of whether we deal with the infinitesimal
canonical transformations of ordinary classical dynamics or those
of Fermion system ``classical'' dynamics), and it is satisfied in
particular for one-parameter infinitesimal $\delta\vec a$ which
are of the form
\begin{equation}
\delta a_i = -\,{i\over\hbar}(\delta\lambda)
{\partial G\over\partial a_i^*}\, ,
\label{Eq:inGen}
\end{equation}
where $\delta\lambda$ is a real-valued infinitesimal parameter
and $G(\vec a,\vec a^{\, *})$ is a real-valued ``generating function''
whose dimension is that of action divided by the dimension of
$\delta\lambda$.  Because $\delta\lambda$ and $G(\vec a,\vec a^{\, *})$
are real, Eq.~(\ref{Eq:inGen}) implies that
\begin{equation}
\delta a_j^* = {i\over\hbar}(\delta\lambda)
{\partial G\over\partial a_j}\, ,
\label{Eq:inGencc}
\end{equation}
and we thus can readily verify that the last of
Eqs.~(\ref{Eq:inCan}) is satisfied.

{}From Eq.~(\ref{Eq:inGen}) or Eq.~(\ref{Eq:inGencc}) we
obtain the form of the equation which governs any continuous
one-parameter trajectory of sequential infinitesimal canonical
transformations in the complex vector phase space:
\begin{equation}
i\hbar{da_i\over d\lambda} =
{\partial G\over\partial a_i^*} \quad \hbox{or} \quad
-i\hbar{da_i^*\over d\lambda} =
{\partial G\over\partial a_i}\, .
\label{Eq:comTraj}
\end{equation}
In the most general circumstance, $G$ may have an explicit
dependence on $\lambda$, i.e., it may be of the form
$G(\vec a,\vec a^{\, *},\lambda)$.  Bearing in mind the relation
$(\vec q,\vec p) = (q_s(\vec a + \vec a^{\, *}),
-i\hbar (\vec a - \vec a^{\, *})/q_s)/\sqrt{2}$ between the
complex and real phase space vectors, Eq.~(\ref{Eq:comTraj})
may be rewritten as the pair of real equations:
\begin{equation}
{dq_i\over d\lambda} =
{\partial G\over\partial p_i}\, , \quad
{dp_i\over d\lambda} =
-\, {\partial G\over\partial q_i}\, ,
\label{Eq:reTraj}
\end{equation}
which are the familiar generalized Hamilton's equations
\cite{Guillemin} that govern continuous one-parameter
trajectories of sequential infinitesimal canonical
transformations in the real $(\vec q,\vec p)$ vector phase
space.

For the case of ordinary classical dynamics (for which the
value of $\mp = +$ in Eqs.~(\ref{Eq:inCan})), the first two
of Eqs.~(\ref{Eq:inCan}) are satisfied identically for the
one-parameter infinitesimal $\delta\vec a$ of the form given
by Eqs.~(\ref{Eq:inGen}) and (\ref{Eq:inGencc}).  However,
for the case of Fermion system ``classical'' dynamics (for
which the value of $\mp = -$), the first two of
Eqs.~(\ref{Eq:inCan}) impose the following constraint on the
real-valued ``generating functions''
$G(\vec a,\vec a^{\, *},\lambda)$ of the continuous one-parameter
canonical transformation trajectories:
\begin{equation}
{\partial^2 G\over\partial a_i\partial a_j} = 0 =
{\partial^2 G\over\partial a_i^*\partial a_j^*}\, .
\label{Eq:Gencon}
\end{equation}
Thus the ``generating functions'' of the continuous
one-parameter trajectories of sequential infinitesimal
canonical transformations in Fermion system ``classical''
dynamics are constrained to be constant or linear in
each of $\vec a$ and $\vec a^{\, *}$, as well as real-valued.
The most general form for such a ``classical'' Fermion
system ``generating function'' is therefore
\begin{equation}
G(\vec a,\vec a^{\, *},\lambda) = G_0(\lambda) +
\sum_k\left (g_k(\lambda)a_k^* + g_k^*(\lambda)a_k\right ) +
\sum_{lm}G_{lm}(\lambda)a_l^*a_m,
\label{Eq:FerGen}
\end{equation}
where $G_0(\lambda)$ is real and $G_{lm}(\lambda)$ is a
Hermitian matrix.  Upon putting this constrained form
for $G$ into Eq.~(\ref{Eq:comTraj}) for the continuous
one-parameter trajectory of sequential infinitesimal
canonical transformations which $G$ generates, we arrive at
\begin{equation}
i\hbar{da_i\over d\lambda} =
g_i(\lambda) + \sum_j G_{ij}(\lambda)a_j,
\label{Eq:iSchro}
\end{equation}
which is a (possibly) inhomogeneous linear equation of matrix
Schr\"odinger form.  (If the $g_i(\lambda) = 0$, this is a
general homogeneous type of Schr\"odinger equation, whereas
if the
$g_i(\lambda) = \hbar\delta_{ik}\delta(\lambda - \lambda')$,
it is a general propagator type of Schr\"odinger equation.)
Thus the ``classical'' dynamics of Fermion systems must be
linear and describable by a Schr\"odinger type of equation.

The generating functions of the continuous one-parameter
canonical transformation trajectories are usually considered
to be {\em observables\/} of classical theory when they
have no explicit dependence on the parameter.  Thus the most
general ``observable'' of Fermion system ``classical''
dynamics must have the form of $G$ in Eq.~(\ref{Eq:FerGen}),
but with $G_0$, $g_k$, and $G_{lm}$ having no
$\lambda$-dependence.  However, when this ``classical''
Fermion theory is quantized by passing (with $\pm = +$)
from the ``symmetric'' Poisson bracket relations of
Eqs.~(\ref{Eq:asPoisrl}) to the anticommutation relations
of Eqs.~(\ref{Eq:asComrl}), it often happens
(particularly in local field theories) that the 
``inhomogeneous'' $\sum_k(g_ka_k^* + g_k^*a_k)$ term of an
``observable'' $G$ is not really, in fact, a bona fide
observable.  Even at the present ``classical'' level it is
always possible to effectively suppress this ``inhomogeneous''
part of an ``observable'' if the Hermitian matrix $G_{lm}$ is
not singular.  This is done by making the canonical
transformation
\begin{equation}
a_i\to A_i = a_i + \sum_j\left (G^{-1}\right )_{ij}g_j.
\label{Eq:conCan}
\end{equation}
It is easily verified that the transformed $A_i$ of
Eq.~(\ref{Eq:conCan}) also satisfy the ``symmetric''
Poisson bracket relations (with $\pm = +$) of
Eqs.~(\ref{Eq:asPoisrl}).  In terms of these $A_i$,
Eq.~(\ref{Eq:iSchro}), specialized to ``observables'',
becomes
\begin{equation}
i\hbar{dA_i\over d\lambda} =
\sum_j G_{ij}A_j,
\label{Eq:hSchro}
\end{equation}
which is of homogeneous Schr\"odinger equation form,
while Eq.~(\ref{Eq:FerGen}), specialized to
``observables'', becomes
\begin{equation}
G(\vec A,\vec A^*) = G_0 -
\sum_{lm}\left (G^{-1}\right )_{lm}g_l^*g_m +
\sum_{lm}G_{lm}A_l^*A_m,
\label{Eq:hFerGen}
\end{equation}
which has no ``inhomogeneous'' term.

The Dirac equation, which is of Schr\"odinger type, can of
course describe a ``classical'' Fermion system, but the
Klein-Gordon and Maxwell equations, although they are
linear, turn out not to be of Schr\"odinger type.  For
example, in one spatial dimension a discretized version
of the Klein-Gordon equation is
\begin{equation}
\ddot q_i -
(c/(2\Delta x))^2
(q_{i + 2} - 2q_i + q_{i - 2}) +
(mc^2/\hbar)^2 q_i
= 0.
\label{Eq:KG}
\end{equation}
This can be replaced by the first-order equation pair
\begin{equation}
\dot q_i = p_i, \quad \dot p_i =
(c/(2\Delta x))^2
(q_{i + 2} - 2q_i + q_{i - 2}) -
(mc^2/\hbar)^2 q_i,
\label{Eq:KGfop}
\end{equation}
which is a version of Hamilton's equations for the
particular Hamiltonian (time evolution generating
function and observable)
\begin{equation}
H(\vec q,\vec p) = {1\over 2}\sum_k\left (p_k^2 +
(c/(2\Delta x))^2
(q_{k + 1} - q_{k - 1})^2 +
(mc^2/\hbar)^2 q_k^2\right ).
\label{Eq:KGHam}
\end{equation}
The constraint given by Eqs.~(\ref{Eq:Gencon})
on Fermion system ``classical'' generating functions
$G$ in the complex vector phase space translates
in the real $(\vec q,\vec p)$ vector phase space
into the two real-valued constraint equations:
\begin{equation}
q_s^2{\partial^2 G\over\partial q_i\partial q_j} =
\left ({\hbar\over q_s}\right )^2
{\partial^2 G\over\partial p_i\partial p_j}\, , \quad
{\partial^2 G\over\partial q_i\partial p_j} =
-\, {\partial^2 G\over\partial q_j\partial p_i}\, ,
\label{Eq:Genconre}
\end{equation}
where the scale factor $q_s$ is real and nonzero.
For the discretized Klein-Gordon Hamiltonian of
Eq.~(\ref{Eq:KGHam}) we have that
\begin{equation}
q_s^2{\partial^2 H\over\partial q_i\partial q_{i + 2}} =
-(cq_s/(2\Delta x))^2 \neq 0
\quad\hbox{and}\quad
{\partial^2 H\over\partial p_i\partial p_{i + 2}} = 0,
\label{Eq:KGHamrl}
\end{equation}
which is {\em not} in accord with the constraint on
``classical'' Fermion system generating functions
that is given by the first of Eqs.~(\ref{Eq:Genconre}).
Thus the Klein-Gordon equation is not of Schr\"odinger
type and cannot describe a ``classical'' Fermion
system.

It is quite clear as well that the old ``four-Fermion''
theory of beta decay is inherently nonlinear and thus
cannot describe a ``classical'' Fermion system (there
is no such objection under the Grassmann variable
regime).

\subsection*{Acknowledgment}

The author wishes to thank T. Garavaglia and S.K. Dutt
for discussions and literature references.

\newpage


\end{document}